# Recent Developments with the licas System

Kieran Greer, Distributed Computing Systems, Belfast, UK.
http://distributedcomputingsystems.co.uk
Version 1.2

*Abstract* - This paper describes recent developments with the licas (lightweight Internet-based communication for autonomic services) software package. In particular, it describes how the architecture and functionality have changed from the first version release. The autonomous nature of the system is focused on, which requires independent behaviour and metadata descriptions of each service. The system has now also been ported to the Java mobile environment. Then some open questions or problems will be discussed in the areas of metadata consistency, security and trust. Finally, some solutions to these problems will also be suggested.

## 1    Introduction

This paper describes recent developments in a lightweight software package that can be used to build distributed service-oriented systems. This system has been described previously [8], but since then it has been modified and extended to include new features. In particular, the architectural changes will be described in some detail. To summarise, the system is called 'licas', which stands for lightweight (Internet-based) communication for autonomic services. It is a software package written in the Java programming language and allows a user to build distributed service-based networks that can also self-organise/self-optimise. Functionality is provided to allow for XML-RPC based message passing and dynamic linking between services. The framework is also suitable for a mobile environment. The architecture and adaptive capabilities through dynamic linking add something new to what other similar systems provide. An open source version of the software can be downloaded from the SourceForge.net web site [14]. Some of the key modifications since the first release of this software are: changes in the architecture, the addition of XML-based metadata descriptors, both to describe and initialise the loaded services, the addition of some real autonomic functionality, and a generic GUI that can be used to test the system. Also, many bugs have been removed and code re-written, to improve security or flexibility.

The system uses an XML-RPC message passing mechanism by default, but it is now also able to dynamically invoke calls on Web Services [11], by parsing WSDL documents and using SOAP. The system is now also compatible with the Java mobile J2ME platform.

The rest of the paper is organised as follows: Section 2 describes some technologies that could be considered to be related to this system. Section 3 describes the new architecture that is being proposed. Section 4 describes the metadata that is now included as default with the system. Section 5 describes some basic autonomic functionality that can be performed. Section 6 gives a description of an admin GUI that can be used to test the system. Section 7 discusses some outstanding questions or problems relating to metadata consistency and security issues. Section 8 gives one example of how the licas system might be used, while section 9 gives some conclusions on the work.

## 2   Related Technologies

There are some recent projects that have aims similar to what licas and its related research would like to do. Two new major projects illustrate how this research might be relevant to future systems. IBM have announced that they want to build computers with 'brain-like' properties [6]. There is also a theory called 'Cloud Computing' [3][4][18]. An earlier Wikipedia-related Web page has described:

'Cloud computing is Internet-based ("cloud") development and use of computer technology ("computing"). It is a style of computing in which IT-related capabilities are provided 'as a service', allowing users to access technology-enabled services from the Internet ('in the cloud' without knowledge of, expertise with, or control over the technology infrastructure that supports them.' This architecture tries to use technologies such as autonomous systems, service-based/grid-related systems, together with abstract or evolving systems that can manage themselves. These could use bio-inspired technologies to self-organise.'

The technologies of service-based, autonomous, bio-inspired and evolving networks are also looked at as part of the licas system research. The definition of cloud computing is also changing and still evolving. The paper [18] gives a useful description of how the architecture has evolved and it appears to have been scaled down now into more practical components that can actually be implemented, rather than purely as a research aspiration. So the more ambitious notions of real autonomy or intelligence are not part of the platforms that are currently being defined, on which the cloud computing systems can be built. There is also quite a strong financial component, where self-configuring services are made readily available over the internet for a particular cost. The idea seems to be to move the computing problem from the computer itself to the internet, in a kind of 'cloud' structure, made of many different distributed and virtual components that can perform many different tasks. Licas is essentially a framework for building service-based systems that can communicate in a distributed manner. It also tries to accommodate the ideas of autonomous behaviour and intelligence. The need for these is also noted in [13], where they argue that in a massively distributed autonomous environment, with a large number of sensors and effectors, manual management is not practical. They also note that a goal-based strategy can be used for self-management. The strategy can create an independent and intelligent knowledge 'cloud infrastructure' for future distributed systems. By goals they mean that agents can use the replies they receive and their state information to achieve their objectives, through their particular capabilities and local knowledge. At the same time, a dynamic knowledge-base with semantic-based negotiations can create a scalable and adaptable learning environment. Such evolutionary environments are critical requirements for the future generation of interwoven computing systems.

Web Services are now part of the licas system, in the sense that you can now call a Web Service from the system components. This call can be dynamically constructed at run-time, by parsing the related WSDL document. The main communication mechanism in licas however is an XML-based RPC (Remote Procedure Call) mechanism. The paper [11] also tries to tie in the Web Services interface with the SOA architecture, while [12] looks at how Web Services can help mobile devices access an SOA. There have been requests to compare the licas XML-RPC mechanism with Web Services. Web Services are an established standard and as licas uses essentially Java Reflection, with some additional metadata, it is probably not productive to try and make a detailed comparison. This is not a new message-passing technique. However, there are some features of the licas message-passing mechanism that might be of interest. Web Services use SOAP as their transport protocol, which is constructed from WSDL. These are highly structured and standardised specifications. They are also quite heavyweight and can even have different port interfaces for different kinds of transport protocol (SOAP, HTTP, etc.). They are only now becoming available in the mobile environment. Licas uses Java Reflection to create a method description directly from the service class itself. This does not require any extra knowledge or description. Through additional metadata, licas also provides a built-in security system with different access levels and there is also the option to send a data value either by parsing it into XML, or by serializing it. SOAP however might also be able to send serialized objects through attachments or string-based descriptions, even though this is not the intention. The licas specification is also more lightweight and already suitable for the mobile environment. Web Services will eventually be made readily available for the mobile environment as well, but licas will probably provide a more lightweight solution. Built with XML and the Java platform, there should not be any problems or conflicts with different kinds of vendor, as is the case for Web Services as well. There are also no problems constructing complex objects in XML-RPC, as you provide your own parser, while there still seems to be a problem with constructing dynamic Web Service method calls. Most method interfaces ask for parameter structures that are only one or two levels deep. One final point to note is the fact that a call in licas can be split into a number of smaller packets and sent using several remote calls instead of just one. The packet size can be specified, when each message is then not larger than this size. This means that very large messages can be sent in parts, making it less likely that a very large message will become corrupted during transmission.

## 3   The New licas Architecture

The aim of licas is to provide a lightweight framework, based on a peer-to-peer (p2p) architecture that can be used for building networks of autonomic service-based components. This could reflect a lightweight Service-Oriented Architecture (SOA) [15]. An HTTP server runs on a machine that can store the network components. These consist of groups of related or nested services. One service can store other services nested inside of it, or it can have permanent links to other services. So a hierarchical structure of nested and/or linked services can be built up. Inside of a single network, each service can communicate locally with any other service. To communicate remotely, a service calls the server that hosts the service to call. The remote server then passes the message to the service being called and invokes that service's method. So any service can become a client, while only the base server classes act as a receiver that can accept remote calls. Figure 1 is a schematic example of the sort of network that can be built.

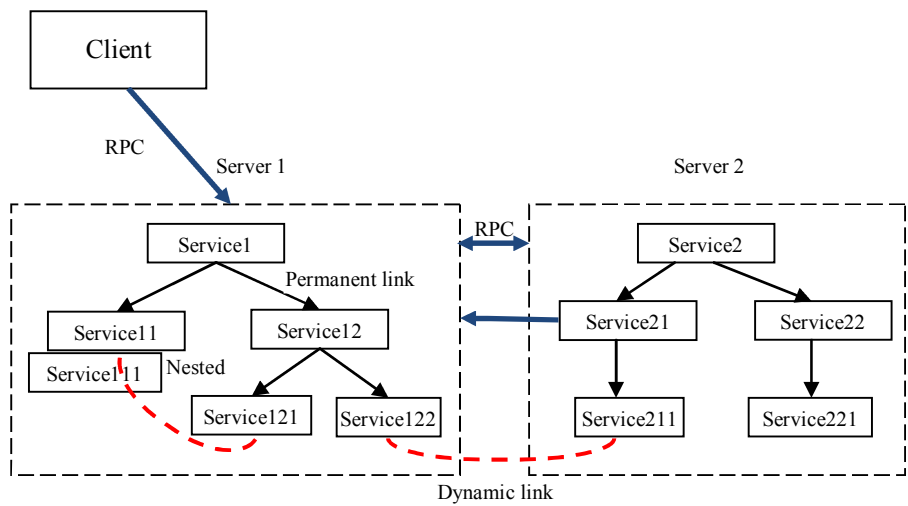

Figure 1.  Basic network architecture

The licas system allows services to be linked to each other now through four different kinds of link.

- **Services related through nesting** – These are links between parent services and services that are nested inside of them. With this association there is not actually any physical link, but one service is stored and directly referenced inside of another. These services must therefore also be in the same network on the same server.
- **Permanent links** – These are permanent links between services that are not directly referenced. The link is defined by a path in one service describing the location of the other service. The permanent links should be used to make up the permanent network structure and so should probably also be limited to services of a single network on a single server.

- **Dynamic links** – These are links that are created dynamically through the use of the system. Unlike the permanent links, these links should be allowed between services on different networks. This is because they not only represent a network structure, but also the dynamic associations generated through its use and these associations can easily span different servers or networks.

- **Service Associations**: These might not be called links, but it is possible simply to store a list of URIs in one service that relate to other services. Permanent links are currently defined as local only, while dynamic links build up autonomously over time, so if you want simply to declare an association between two services in a single call, you can add the URI of one service to a structure in the other service. As a link is not a physical object, these can also be called links, depending on whether they make up the general architecture or not.

## 4   Metadata and Administrative Features

Metadata is 'data about data' and can be used to describe the contents or functionality of a system. The Semantic Web [1] for example will use metadata to describe the contents of Web pages, so that programs can automatically understand what the content is about. The licas system now provides capabilities for adding and using default metadata objects, so that the functionality of the services loaded into the system can be described. The metadata is described by XML and needs to take account of the different types of service or link that is being represented. Metadata should be stored for each individual service. However, if the service is a common or utility service, then possibly only one permanent set of metadata should be stored for all of its instances. For example, if we add linking services to lots of other services, the linking service should have only one set of metadata that does not change and would therefore define that service type exactly. Each service can generate a certain amount of metadata automatically, from the information that is used to create it. A skeleton structure of this metadata is shown Figure 2.

```xml
<?xml version="1.0" encoding="UTF-8"?>
<xs:schema elementFormDefault="qualified" xmlns:xs="http://www.w3.org/2001/XMLSchema">
    <xs:element name="Service_Meta">
        <xs:complexType>
            <xs:sequence>
                <xs:element maxOccurs="1" minOccurs="1" name="Service_Type" type="xs:string"/>
                <xs:element maxOccurs="1" minOccurs="1" name="Description" type="xs:ENTITY"/>
                <xs:element maxOccurs="1" minOccurs="1" name="Other_Meta" type="xs:ENTITY"/>
                <xs:element minOccurs="1" name="Class_Name" type="xs:string"/>
                <xs:element minOccurs="1" name="Handle">
                    <xs:complexType>
                        <xs:sequence>
                            <xs:element name="U" type="xs:string"/>
                            <xs:element maxOccurs="unbounded" minOccurs="0" name="S" type="xs:string"/>
                        </xs:sequence>
                    </xs:complexType>
                </xs:element>
                <xs:element maxOccurs="unbounded" minOccurs="0" name="Jar_File" type="xs:string"/>
                <xs:element minOccurs="0" name="Constructors">
                </xs:element>
                <xs:element minOccurs="0" name="Methods">
                </xs:element>
                <xs:element minOccurs="0" name="Child_Service_Meta">
                </xs:element>
                <xs:element minOccurs="0" name="Link_Service_Meta">
                </xs:element>
            </xs:sequence>
            <xs:attribute name="uuid" type="xs:string" use="optional"/>
        </xs:complexType>
    </xs:element>
</xs:schema>
```

Figure 2. Skeleton structure of an internal metadata document that can be used to describe a service

Common information that is stored includes describing the service type and a general description of what the service does. The type can be represented by a string; for example, 'query engine', 'metadata processor', etc. A general description should also be stored, which is an XML description of what the service does. This element can then be retrieved and read to define the general functionality of the service. Also required is the URI location of the service on the server, which is defined as the service path description. From this the user will know where the service is located and thus how to find and call it. For example, a path description to a service called 'Service2' that is nested inside of a service called 'Service1' could look like:

<U>http://1234.5.6.7:8888</U><S>Service1</S><S>Service2</S>

The interface to the service would also include descriptions of the publically available methods, describing their parameter specifications. If you want to create a new similar service from the metadata description, then you need additional information, such as what class the service is and also the constructor definition. If the class needs to be loaded remotely, then a list of jar file URIs also need to be provided. While the service can generate a certain amount of metadata automatically from the information used to create it, each service can also then be initialised with additional metadata from an admin document. This information can define other descriptive or security features of the service. This document should return similar information to what the service would generate itself and is described fully in the 'licasAdminGuide' document that is part of the licas download package.

In addition to default fields, such as the service type, or service description; the admin document can also provide the following information: The current system can be quite open, where if you have access to a service, you can have access to any of its methods. So to provide for better security, it is possible to partition the methods into groups with differing levels of security. This is particularly important for autonomous systems, as they need to be able to independently negotiate with each other, to determine if one component will allow another component to use its operations. In autonomous systems this is done through 'Service Level Agreements', which are essentially contracts agreed between two parties, for one to provide a service to the other. If two services agree on a contract for one particular transaction however, the server service would not want to allow the client service to perform other transactions as well. To try and help with this problem, the service methods can be grouped into different security or access levels, each with their own password. The access levels can also be put into mutually inclusive or exclusive groups, where one password for a higher level group would also allow access to all groups at lower levels, except for the excluded ones. The alternative is probably a unique password for each transaction that is carried out on any method, which would probably be linked to a financial payment. The idea of access levels is to have a more free or open system, but with added security. You would only need to trust the program using your services to allow it unlimited access to certain ones. The admin document also allows you to specify classes for loading that can be used as part of the Autonomic Manager system, for monitoring the service autonomously. Note that the system only provides the framework for specifying the access levels. The autonomic services must decide on contracts between themselves and this not only requires the metadata descriptions, but also the intelligence to understand these and also to negotiate over them. Any intelligence that is required would need to be programmed into each service and could possibly be different for each individual service. The process would be very much like the negotiation protocols of agent-based systems.

There are two additional fields that the administrator can use to initialise the service with extra information. The first field can be used to allow an administrator to add any additional metadata descriptions that might be specific to the service being loaded, as well as the default description. The second field allows the administrator to initialise a service with some real data. This is simply passed to the service as XML in the specified field. The licas base classes do not have any specific structure for storing or manipulating data themselves and so this field only provides a place from where data can be retrieved by a service that is derived from one of these base classes.

## 5 Autonomic Functionality

A network constructed with licas would show autonomic functionality, through allowing the services to self-organise. A dynamic linking mechanism is included with the software package to show how the services can do this. The services can form temporary links with other services that they are typically associated with. These links can be created or destroyed depending on the system use and so reflect the current state of the system. This has been written about in detail in previous papers, for example [9] and test results suggest that when dynamic links are used, an 80-90% reduction in search might be achieved with only a 5-10% reduction in the quality of answer. More recent tests have resulted in slightly different values under slightly different criteria, but still very good. It is also possible to create permanent links between services, which can also be done autonomically, by adding behaviours to the services. One of the default classes is an 'Auto' class. This is intended to provide for more agent-like or autonomous functionality and now has a default implementation for its Thread's 'run' method. If the thread is run, then the class will try to load and run a behaviour class. The behaviour will try to load in an evaluation function and use this to evaluate associations with other services. The behaviour, for example, could ask other services for values relating to something and depending on the reply decide whether to link with them or not. The behaviour and evaluation processes can be very simple or much more complex, realising a wide range of autonomous capabilities.

## 6   Default GUI

The licas system now also provides a GUI that can be used to test some of its features. This GUI can also be used to test self-organising algorithms or even query the system. Figure 3 shows one of the GUI panels with the related network running.

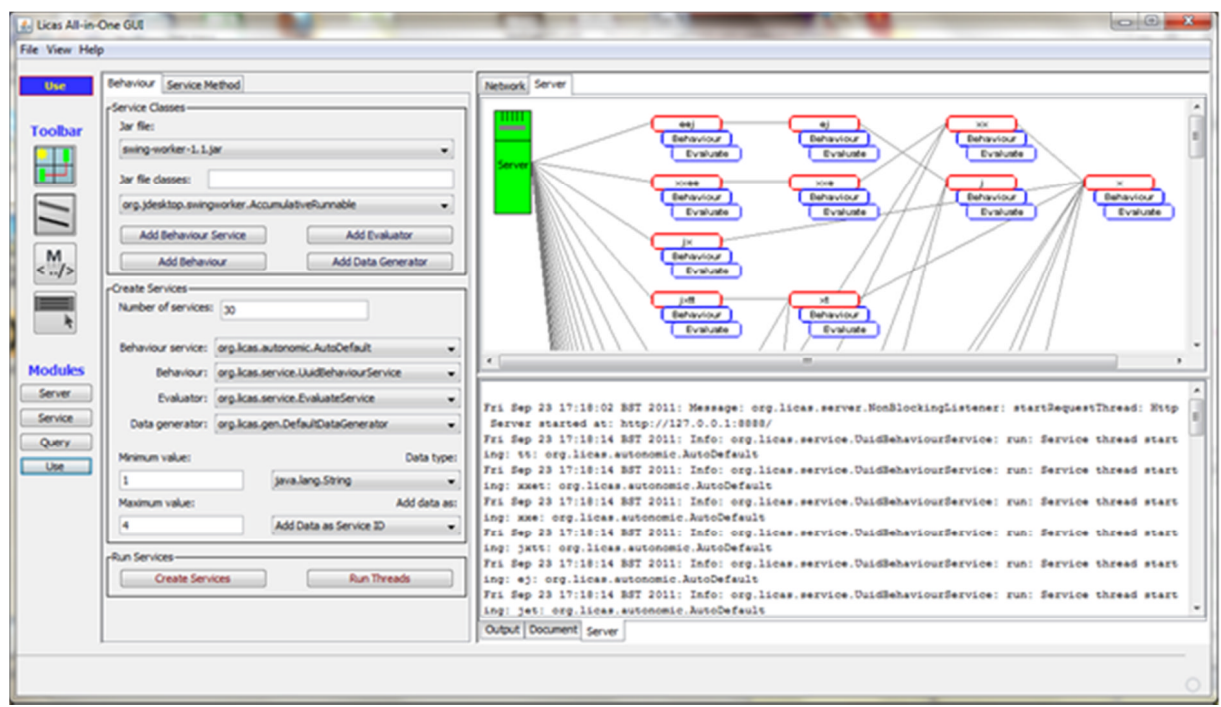

Figure 3. Admin GUI running test behaviours for self-organisation

This panel allows the user to test his own algorithms, where each service will try to self-organise using some pre-defined behaviour. The GUI options are flexible, so that the user can load in his own specific classes, or use the default demonstration ones. The user can therefore create a number of services based on the selected classes and run their threads to see how the network would autonomously self-organise. In the default example, the classes generate string-based values of a specified length. These are then assigned as the IDs of the services. The services then self-organise based on the relative values of their IDs. While the algorithm provided is only intended to illustrate the functionality, it does show that the services are capable of autonomous behaviour, no matter how simple this might be. A more complex behaviour could be run in the same way.

The GUI is intended to be generic and only provides a view of the network, which is generated through the metadata retrieved from the network. The text box on the bottom right shows some metadata that has been retrieved. The metadata object and parser are standard and supplied by the package, but the service class type can be any class and so any service can be loaded into a network and then displayed on the GUI. Other panels allow you to manually load in services from non-local jar files. There is also a popup menu when you right-click one of the services on the display that allows you to manually create or destroy permanent links. The dynamic links can span different networks and so it is not easy to represent these on the network graphic. However, the popup menu allows you to retrieve some of the dynamic link information for a selected service as well. There is also the option to expand or close specific nodes, or even display the network to only a certain depth, so that it can be shown on the graphic. The network graphic itself runs on a separate thread that automatically updates itself every second or so. So a current view of the network is always available. The metadata will also automatically update itself when the service information changes.

# 7  Open Questions Regarding Metadata and Security

Generating metadata documents has created a number of open questions about how to make this metadata consistent over the whole system. For example, the current implementation assumes that if a service has the same ID as another one, then it should also have the same metadata document. It might be necessary to give different services the same ID, because if one linking service wants to call another one for example, it needs to know that it is calling another service that has a 'link' ID. If this is not the case, then it is much more difficult to program this sort of activity. The current architecture would therefore suggest that similar utility services would all have the same metadata, but other individual services could have their own metadata objects that would dynamically change with the system use. This consistency would also mean that the user of the system would know that these utility services would all behave in the same way before using them. However, it also means that these services cannot themselves really store any dynamic information that might change over time. For example, the utility services could not store any dynamic links. They could be used to create or even store dynamic links for a parent service, but could not have any dynamic link metadata themselves. This is not as flexible as allowing all services to dynamically change their metadata, but provides some level of standardisation. Maybe the following rule would be a good one to follow:

*Same service ID – same service type – same service metadata*.

It would also be important to consider the fact that some metadata can be private while the rest is public. In that case it might only be the public metadata that would need to be static and consistent, while the private metadata could dynamically change for any service. Table 1 gives a description of the different scenarios where metadata would be available. It compares unique services or utility ones, static or dynamic metadata and public or private metadata; stored for the service in question, or its parent service. The table shows that the only problem would in fact be to store dynamic public metadata for a utility service. It has been argued that the Internet is losing its end-to-end, peer-to-peer nature [5]. While it is still needed, hosts can no longer act both as a server and client, because of issues of trust, security and authentication. Trust can be defined as knowing that whoever you are communicating with is a legitimate peer in a transaction. Authentication is ensuring that you are who you say you are and security is knowing that a properly authenticated connection from a trusted peer or user is secure from attacks.

| Metadata | | | | | | | | | | | | | | | |
|---|---|---|---|---|---|---|---|---|---|---|---|---|---|---|---|
| Unique Service ID | | | | | | | | Shared Service ID | | | | | | | |
| Static | | | | Dynamic | | | | Static | | | | Dynamic | | | |
| Public | | Private | | Public | | Private | | Public | | Private | | Public | | Private | |
| This | Other | This | Other | This | Other | This | Other | This | Other | This | Other | This | Other | This | Other |
| Yes | Yes | Yes | Yes | Yes | Yes | Yes | Yes | Yes | Yes | Yes | Yes | No | Yes | Yes | Yes |

Table 1. Table describing the relative flexibility of metadata for different types of service. A 'yes' in the final row means that metadata is available for that combination and a 'no' means that it is not.

The paper just referenced suggests solutions to this problem and the licas architecture would also be helpful in this respect. It is more of a hybrid architecture, where every call firstly goes through a base server before being passed to the service in question. Each service is not part of a pure p2p architecture and while it can send messages and act as a receiver also, the server aspect is placed centrally in a base server. The base server could be provided by a more established partner that would allow users to then load in their own individual services. This unit could thus perform certain security checks that each service loaded into the network would then not have to provide. For example, security against attacks of some sort. The access levels would also be helpful in this respect.

If we focus on a truly autonomous system however, then trust and authentication are more difficult. The services need to be able to determine these for themselves, but how can this be done? How can you tell that a service that wants to negotiate something is genuine? One solution is to have each service that operates, register itself with some sort of central registry. The service can then give the calling service its registration key, the name of the company that wrote it and the registry on which it is registered. If these credentials can be verified, then the service can be considered as genuine. This might work for something like a Microsoft service, however, with any number of smaller companies providing services, you could easily find a service belonging to a company that you were not familiar with and then not be sure if it was genuine. If these services were simply refused, then this would provide a real problem with regard to competition. One other possibility would be to have a broker or mediator that could mediate the transaction between the two services. This mediator would have to be known and established beforehand, where the scenario could be as follows:

One client requiring a service finds a server that can provide the service at a certain cost. The two services agree to the transaction but then also agree to use a mediator to help with the transaction. The client gives its monetary details to the mediator and then the mediator asks the server to carry out the service. The server carries out the service and sends the result to the mediator which returns it to the client, or directly from server to client. If the client is happy with the result, then the mediator releases the monetary details to the server and the transaction is complete. If the client is not happy with the result, then the server and the mediator need to determine that the service reply was genuine. This could be a case of the mediator verifying that the service provider is genuine. If this cannot be completely verified, then possibly human intervention would be required between the mediator and the service provider's company, until the service could be proven to be genuine. If the service is shown to be genuine, then this information also needs to be sent to the client and the monetary matters completed. The main advantage of this is that there is a block or check, so that a transaction will not automatically take place between the two autonomous components. If the transaction is genuine however, then it is also guaranteed to go through. While this could eventually end up with human intervention, at least the check would be in place and if most services were genuine, meaning that disputed transactions would be relatively rare, then this could provide a certain level of security against any bogus ones. Encryption could also be used to protect sensitive information, with keys being passed to the relevant components only. For example, the mediator is given the key value instead of the actual monetary details. Examples of negotiation protocols between services can also be found in [16] and [17]. In particular, the credential verification process is described, but also the use of a mediator, or negotiator, on behalf of the service.

The use of a third-party negotiator however assumes that the client still has some idea of what sort of result it should get for the query request. Also, if the reply is part of a chain of transactions, as in a business process for example, then one service might be corrupt while all of the others would be legitimate. So if there was an error or query over a result, there might be some time delay to determine exactly which service was at fault before any of the service participants could be paid. The author has naively stated previously [7] that what is important about the intelligence in a service is that it is 'good'. This is of course a meaningless statement intended for an idealistic situation, but it is difficult to define this requirement clearly in some way. If we could test this as part of a process, then we would know that the application was providing an appropriate level of service. What if a client application could ask several questions, where it would already know the answer to all but one of them. The service provider would then not know which question was the genuine one and so would have difficulty in cheating with its reply. A kind of game scenario could arise, where the provider would have a 50% chance of getting away with a wrong answer if only two questions were asked, to less than this for more questions. These electronic services would probably be able to process more than one request as part of the process and still make the transaction practical. This would then mean that services could negotiate with each other autonomously and have added levels of confidence that the transaction would be legitimate, without requiring all of the human knowledge and intelligence of business transactions. The problem with these electronic services is that they can easily disappear, but on the other hand, they should not have any prejudice over carrying out a transaction and will only perform what they are programmed to do.

## 8   Example Application

Before a possible application of the system is described, it would be useful to review the linking mechanism that is used to self-organise. The system constructs networks of nodes or services. These nodes can be dynamically linked. The dynamic links are created through the system use and are defined by chains of concepts that describe what the link is for. These concept chains can also be hierarchical in nature, with a base concept that branches out to other ones, and so on. The hierarchy can then be traversed to find linked information relevant to the current search. This is meant to be used as part of something like the Semantic web, where several Web pages need to be associated together to provide a complete answer.

One example of how this system might work would be the travel agent organiser often associated with agent-based systems. In this case the organiser tries to link related sites for different parts of the travel program. The scenario for booking conference travel details could be as follows: A number of visitors have been to the specified location and found the travel arrangements to be good. They upload their travel details under the appropriate concepts. For example:

Conference dates: 1 – 3 December 2009.
Conference location: Country, city and street.

- The person finds a hotel and stores the details under the concepts of 'hotel – country – city – street – dates - cost'.
- The person finds flights and stores the details under the concepts of 'flight - destination_airport - arrival_airport - flight_courier - flight_dates - flight_times - cost'.
- The person finds a connecting train or other transport and stores the details under the concepts of 'connection_transport – country – city – street – type – date – time - cost'.

These details are stored in the dynamic linking structure, where the concepts define the search paths leading to each reference and the linking structure stores the dynamic links between the references. Another person can then search this database and retrieve the linked information. If another person asks for only flights to the specified city from his local airport, then all flights or branches under the concepts of:

'destination_airport - arrival_airport - flight_courier'

can be retrieved. If the user also asks for particular dates and times, then these links can also be traversed, or filtered, instead of all of these branches being retrieved. Finally, the budgetary considerations can also be made. This sort of database or linked list can self-organise when more information is added and used to build reliable links. For example, if the user enters three different pieces of information as indicated above, it is understood that these go together and temporary links between them can be made. These can then be reinforced by further entries until the information is made reliable and can be retrieved as part of a search.

The initial requirement for a search could then be: Travel to France for a conference in Paris. Need a hotel in Paris where the location is near the conference building and transport where the travel is on the days d1 and d2, at times t1 and t2, with budget less than x. This could translate into a search query that looks something like:

Select hotel, flight, connection where (hotel.street equals conference.street) and (hotel.city equals Paris) and (hotel.cost less than £xxx.xx_per_night) and (train.arrival.station equals near conference_street) and (train.departure.station equals near_airport) and (train.departure.time equals flight.arrival.time_plus_x_hours) and (flight.arrival_day equals d1) and (flight.departure_day equals d2) and (flight.departure equals morning) and (flight.cost less than £xxx.xx).

If services were added along with the linked references, then they could calculate distances, where words such as 'near' or 'approx' could be added. They could also notify the user when the travel times changed, or even send a text when good deals were found. More elaborate services could also negotiate or even book the conference, etc, as the agent-based system would do. If travel experiences or ratings were also added, then services could learn what other options were good or bad and take this into account as well, and so on. Another option, as written about in [1], would be to allow links to events or attractions that are close to a hotel, where the visitors would rate the attractions that they visited. This could also be built up in a dynamic way. This is also quite similar to the wiki-style updates suggested for semantic search engines in [10].

# 9  Conclusions

The licas system has been re-written in large parts, to provide a much more stable and robust system. This has also provided the possibility of adding new features, including porting to a mobile environment and Web Service invocation. Metadata is now included as part of the framework and can be used to describe a service and the methods that it provides. The metadata can also provide added security, as methods can be grouped into inclusive or exclusive sets that might require different passwords for access. A test GUI is now available and can be used to demonstrate the autonomous behaviour. The GUI only provides a view of the network and so is generic with regard to the type of service being processed. It would also allow the user to load in his own services, to test his own algorithms. Some problems with truly autonomous behaviour have been written about, in particular, the problems with trust and security for autonomous transactions. Some solutions have also been suggested that could provide the required level of security. This would allow a user to confidently run an autonomous service or system and allow it to perform his/her own important operations.